\documentclass[a4paper]{article}
\usepackage[utf8x]{inputenc}
\usepackage{graphicx}
\usepackage{authblk}
\usepackage{a4wide}
\title{Extension of calibration of UV-VIS spectrometer from VIS to UV region by using single bubble sonoluminesce}

\author{J. Anto\v{s}}
\affil{Institute of Experimental Physics, SAS, Ko\v{s}ice, Slovakia}

\begin{document}
\maketitle
\stepcounter{section} 
\begin{abstract}
A simple method to calibrate UV-VIS spectrometer by means  of single bubble sonoluminescence is presented and tested. 
\end{abstract}

\section{Introduction}
 A sonoluminescence is a phenomenon known for almost a century however is still lacking a generally accepted explanation. After discovery
 of single bubble sonoluminescence \cite{gaitan} even more questions have been risen than provided answers by a possibility to study this 
 phenomenon at well defined conditions.
  Purpose of this paper is to take advantage of known property of Single Bubble Sonoluminescence (SBSL) in water - specifically black-body 
  radiation like shape of spectra (see e.g. \cite{SLbb}).\\
   We assume spectrometer is already calibrated in VIS region 400-900 nm.  This can be done by using relatively inexpensive  tungsten halogen
   standard calibration lamp. We fit observed single bubble sonoluminescence spectrum in region 400-900 nm by a function proportional to
    Planck's distribution and functional dependence of fitted distribution can be used  below 400 nm for correction of response of spectrometer
    - to calibrate it. This is brief idea of proposed procedure. Details will be given in sections below.\\
    Paper is organized as follows. There will be given short description of apparatus to produce SBSL and to observe spectrum of SBSL.
    Then details of fitting procedure and correction function will be given, calibration of spectrometer in region below 400 nm determined.
    Calibration will be tested by a comparison of observed tryptophan  fluorescence spectra with published standard.

  \section{Apparatus}
  
\begin{figure}[h]
\begin{center}
\includegraphics[width=3in]{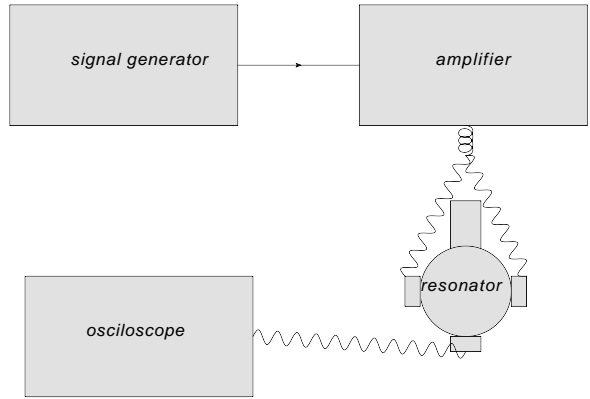} \\

\caption{schematic view of apparatus for SBSL
\label{SLsch}
} 
\end{center}

\end{figure}

  Apparatus consists of equipment to produce SBSL (resonator, amplifier, signal generator - see Fig. \ref{SLsch}) and spectrometer QE65000 \cite{ooqe65} to
  measure spectra. For testing
  of goodness of calibration we used fluorophore tryptophan which is characterized by a peak of excitation spectra around 265 nm and emission spectra
  310 - 450 nm range. Deuterium lamp was used for excitation (without any filter) of tryptophan and 3 way cuvette holder to measure spectra which
  are signal dominated (emission spectra) and separately spectra which are background dominated.\\
   Apparatus for SBSL we use is sort of standard. We will just  emphasize some details below we consider
   important for present analysis.

  \subsection{Resonator}
  Beaver flask component of resonator is standard laboratory glass. We tested couple resonators made by different producers. In this paper results are based
  on  quartz beaver 100 ml flask resonator. \\
  Flask is equipped with two piezoceramic transducers glued there facing each other on equatorial side of beaver flask. Their purpose is to create standing 
  ultrasonic wave which traps gas (air) bubble when flask is filled with proper liquid. At the bottom of beaver flask is glued small piezoceramic 
  element which serves as a microphone. As a working liquid we used water. At a proper conditions (frequency and amplitude set in signal generator) bubble oscillates,  expands and implodes
  and  produces short burst of light per period. 
  \subsection{Interface of optical system and spectrometer}
\begin{figure}[h]
\begin{center}
\includegraphics[width=3in]{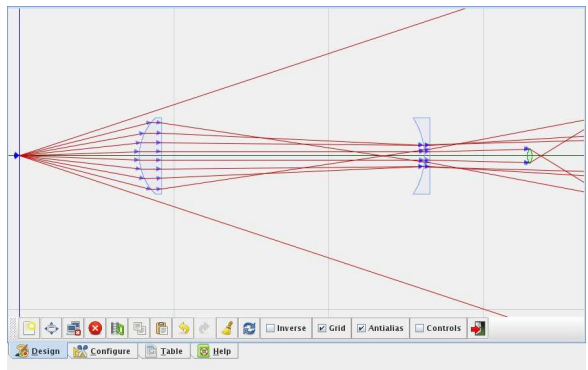} \\

\caption{Optical configuration for measurement of spectra of SBSL
\label{lensestd}
} 
\end{center}

\end{figure}

   Schema of optical configuration is displayed in Fig. \ref{lensestd}. Optical system consists of 3 lenses made from fussed silica with focal length
   $f_{L}$ = 38, -50 and 10 mm respectively.
   The last one is coupled to optical fiber through connector sma905 and other end of a fiber is through the same type of connector connected to a
   spectrometer. System was tuned in such a way that ``shining bubble'' was on optical axes of the above optical system. Optical system is placed
   as close as possible to resonator.
   
  \subsection{Spectrum of SBSL}
\begin{figure}[h]
\includegraphics[width=10cm]{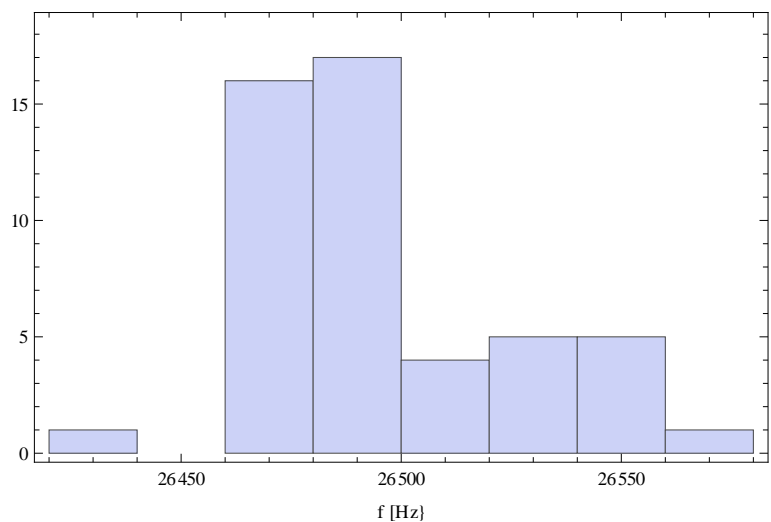} 
\caption{histogram of starting frequencies at which SL was achieved} 
\label{SLfrq}
\end{figure}

\begin{figure}[h]

\includegraphics[width=10cm]{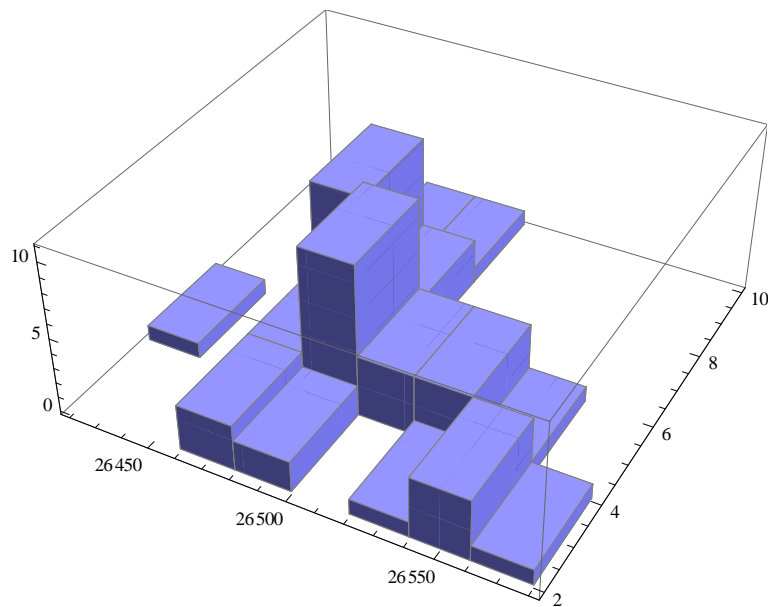} 
\caption{A 3d histogram of starting frequencies x signal amplitudes when SL was achieved} 
\label{freqamp}
\end{figure}

  Among the parameters which have been controlled to achieve SBSL have been frequency (Hz) and amplitude (V peak to peak) of signal generator.
 Distribution of frequencies  for which sonoluminescence was achieved (for a given resonator)  is in Fig.\ref{SLfrq}.
Mean of above distribution is 26.478+-0.03 kHz. 
 A correlation between starting frequency and peak to peak amplitude of signal produced by signal generator is displayed in Fig. \ref{freqamp}.
 
\begin{figure}[h]

\includegraphics[width=10cm]{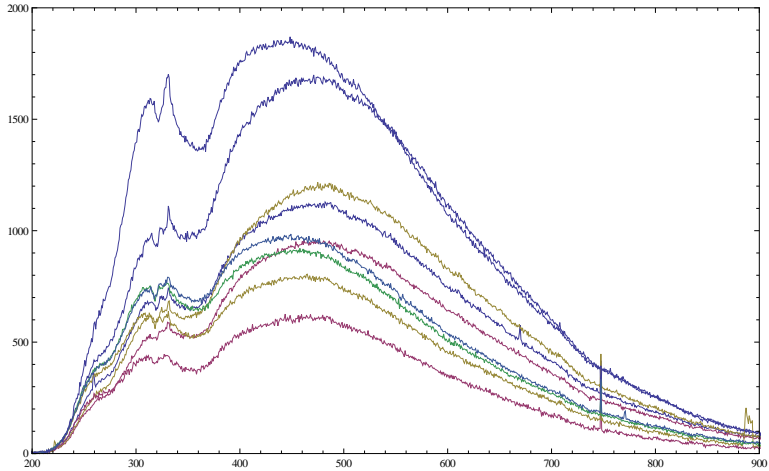} 
\caption{raw dark current subtracted spectrum of SBSL of sample selected for calibration of spectrometer} 
\label{rawsbsl}
\end{figure}

\begin{figure}[h]

\includegraphics[width=10cm]{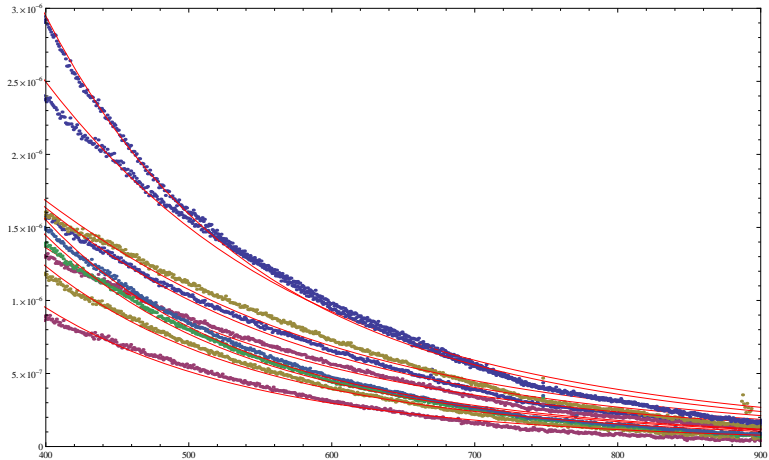} 
\caption{Same sample as in Fig. \ref{rawsbsl} spectra corrected by absolute irradiance calibration of spectrometer in 400 - 900 nm wavelength region. Results of black-body 
fits are included there (red lines).} 
\label{lcalsbsl}
\end{figure}

  Raw (dark subtracted) spectra of SBSL are in Fig. \ref{rawsbsl}. Corrected (calibrated) spectra in region 400 - 900 nm are displayed in Fig. \ref{lcalsbsl}.
    

\section{Fitting procedure and calibration of spectrometer in region below 400 nm}
Black-body radiation like shape of distribution from sonoluminescence can be used for calibrating spectrometer in UV region provided it is 
calibrated in VIS region.\\
 Absolute irradiation calibration of  spectrometer in region of 400 - 900 nm is done by using LS-1-CAL calibration lamp \cite{ools1}. Idea is to
 fit absolutely calibrated spectrum of SL in this region by black body radiation formula and region in ~250 - 400 nm region calibrate
 by assuming that true spectrum in this region is consistent with fit from 400-900 nm region. 
 \subsection{Black-body radiation fit}
  Directly to fit Planck distribution to SL spectrum is problematic. Fit is very unstable and for reasonable outcome it needs a good
  initial starting point. We tried several approaches and one which proved most successful will be described below.\\
  We use form of Planck distribution below
  \begin{equation}
  f_{1}(\lambda,T) =\frac{C_{1}}{\lambda^5} \frac{1}{exp(\frac{C_{2}}{\lambda T})-1}
  \end{equation}
  and also approximate form\\
  \begin{equation}
  f_{2}(\lambda,T) =\frac{C_{1}}{\lambda^5} exp(-\frac{C_{2}}{\lambda T})
  \end{equation}
  where $\lambda$, T is wavelength and temperature respectively, $C_{1}, C_{2}$ are standard constants in Planck distribution. We do not fix
  scale of distribution and leave it as a free parameter. Therefore by fit to SL spectrum we try to determine two parameters a and T of function
  $a f_{1}(\lambda,T)$ (or $a f_{2}(\lambda,T)$.\\
  To determine good initial approximation we define new variables r=ln(a) and z=1/T. We can fit easily r + $ln(f_{2}(\lambda,1/z)$ to natural
  logarithm of the data. This can be done even analytically, because in this case it is linear fit.\\
  Initial values $a_{in}=exp(r_{fit0})$ and $T_{in}=1/z_{fit0}$ are supplied into nonlinear fitting procedure for fit $a f_{1}(\lambda,T)$.
  Fit was done in framework of mathematica package \cite{math}.\\
   
\subsection{Application of fitting procedure to SL data}
  Region 400 - 900 nm can be considered as safely covered by LS-1-CAL lamp calibration. However this calibration does not cover all 
  parts of optical system. There is water in resonator, glass, lenses. At least for a moment we ignore these sources of deformation of 
  true spectra.
 \subsubsection{Calibration of region below 400 nm}
\begin{figure}[h]
\begin{center}$
\begin{array}{cc}
\includegraphics[width=2.5in]{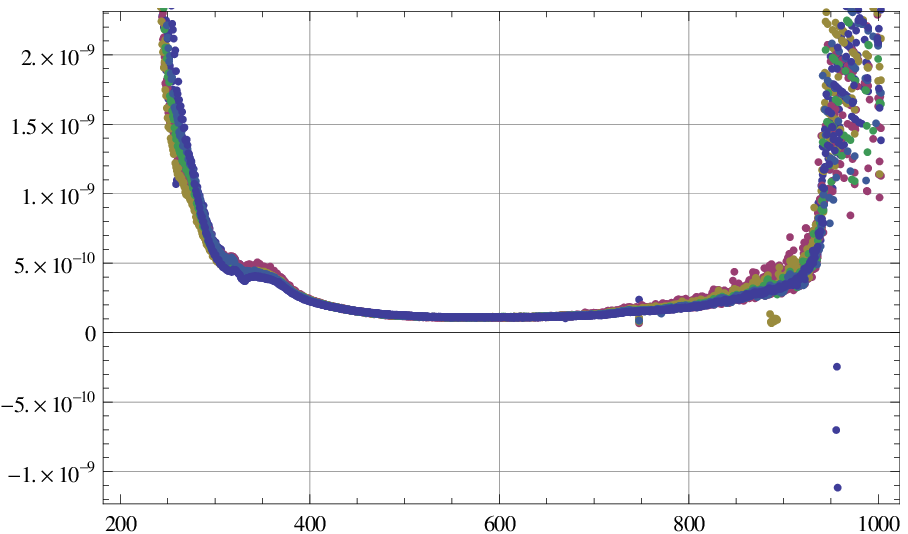} &
\includegraphics[width=2.5in]{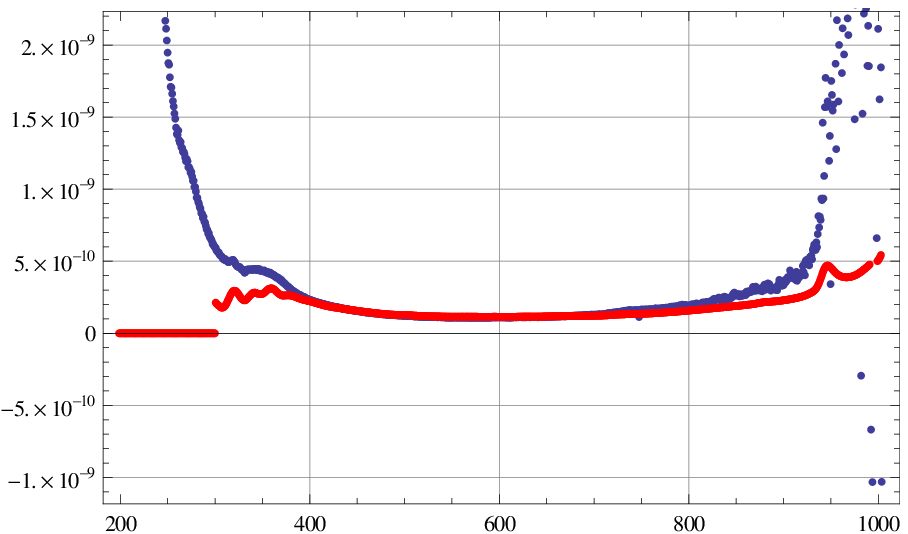} \\
\includegraphics[width=2.5in]{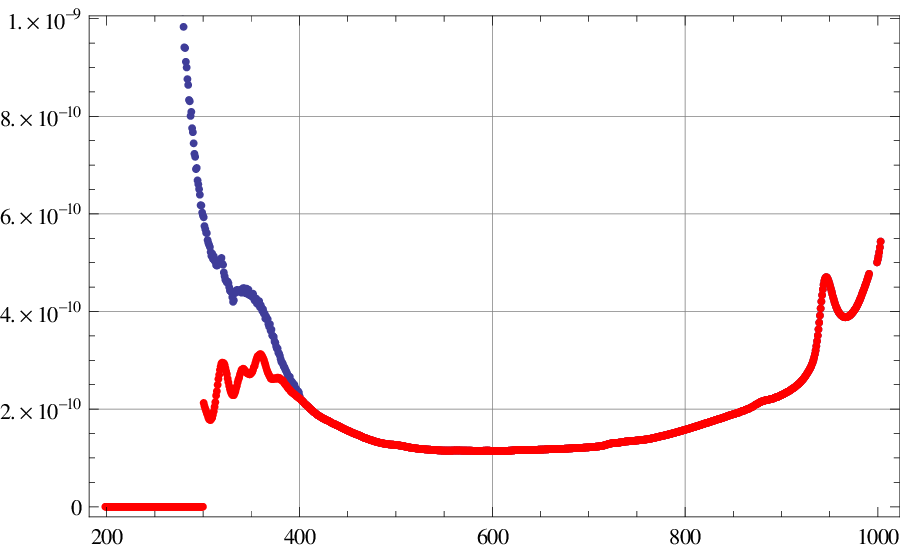} &
\\
\end{array}$
\end{center}
\caption{Upper left - derived calibration functions from selected SBSL sample, upper right - average calibration function (from selected) compared with original
calibration function based on lamp LS-1-CAL, bottom left - combined calibration function.
} 
\label{lconvlcomb}
\end{figure}

  If we assume that Planck's distribution fit determined in region 400 - 900 nm is valid below 400 nm (up to ~250 nm) we can determine
  general calibration function for this spectrometer (and optical system used) up to ~250 nm. Calibration function is characteristics of 
  spectrometer (and optical system used) and not process (SL) used for its determination. Therefore if above assumption is valid
  calibration function determined by SL at different conditions should be very close.
Calibration function is defined
\begin{equation}
 C(\lambda)= \frac{F_{t}(\lambda)  }{ \frac{\delta N}{\delta_\lambda } } \Delta t S
\end{equation}

where $F_{t}(\lambda)$ is known true spectrum  (in absolute units)  of the source, $\frac{\delta N}{\delta \lambda}$ is raw
dark subtracted spectrum of the same source determined by spectrometer, $\Delta t$ - integration time over which spectrum was accumulated, S - 
surface area from which signal was collected.\\
 In our case as a $F_{t}(\lambda)$ is taken Planck's distribution with fitted temperature.  In Fig. \ref{lconvlcomb} one can see evolution of above steps.

As a final calibration function is considered combination of average of derived calibration functions below 400 nm and original calibration
function (based on LS-1-CAL lamp calibration) for wavelength greater than 400 nm.  One can see from Fig. \ref{lconvlcomb} original calibration
in region 300-400 nm is quite different from new one in this region. Fact that original calibration failed in 300-400 nm region was notoriously known to us. 
\clearpage
\subsection{Test of new SL based calibration}
 To test new calibration we decided to measure emission spectra of fluorescence of D-L,L-tryptophan. D-L tryptophan is well known
 fluorophore, with emission spectra well measured and suggested to be used for calibration of fluorometers \cite{gard}. 
 \subsubsection{Measurements done by spectrometer QE65000 with SL calibration extended below 400 nm}
\begin{figure}[h]
\begin{center}
\includegraphics[width=3in]{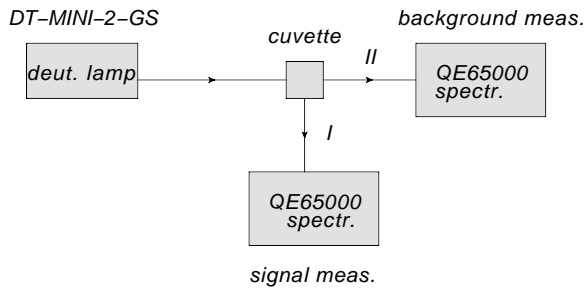} \\

\caption{Configuration for test of calibration by measurements of emission spectra of D-L,L-Tryptophan}
\label{testcal}

\end{center}

\end{figure}

 Our measurement setup consisted of a three way cuvette holder, deuterium lamp (GS-DT-MINI \cite{oodtmini}) and spectrometer QE65000.
 Deuterium lamp was connected via sma905 connectors to cuvette holder and created beam-line (see Fig. \ref{testcal}). 

\begin{figure}[h]
\begin{center}$
\begin{array}{cc}
\includegraphics[width=2.5in]{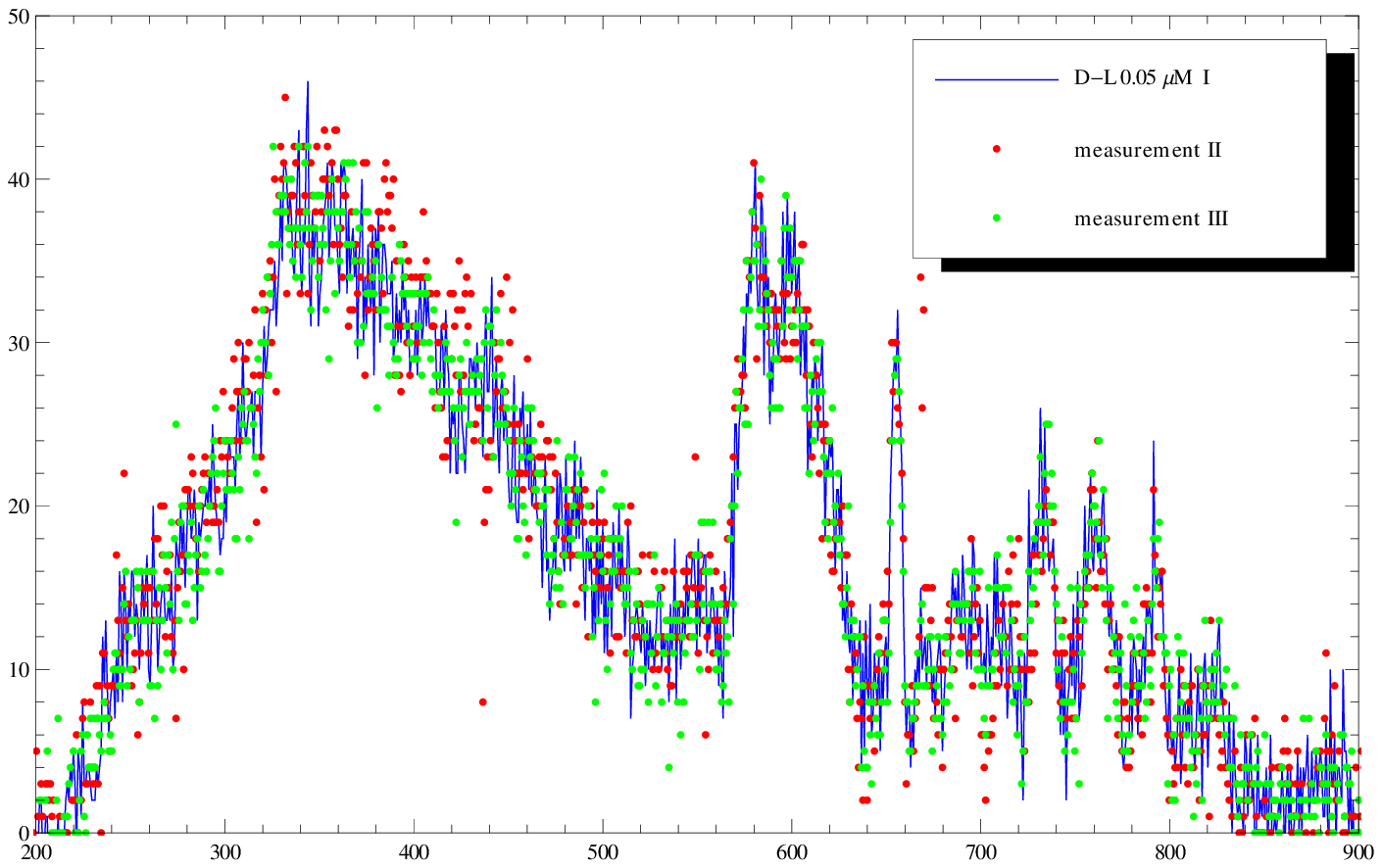} &
\includegraphics[width=2.5in]{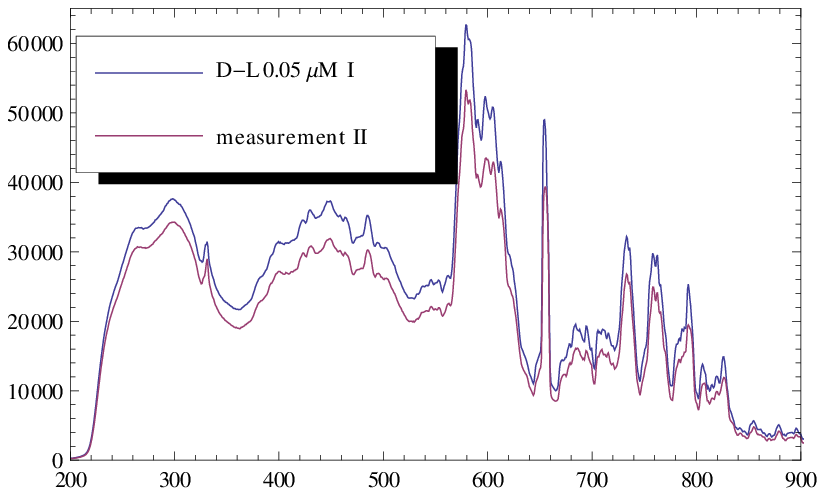} \\
\includegraphics[width=2.5in]{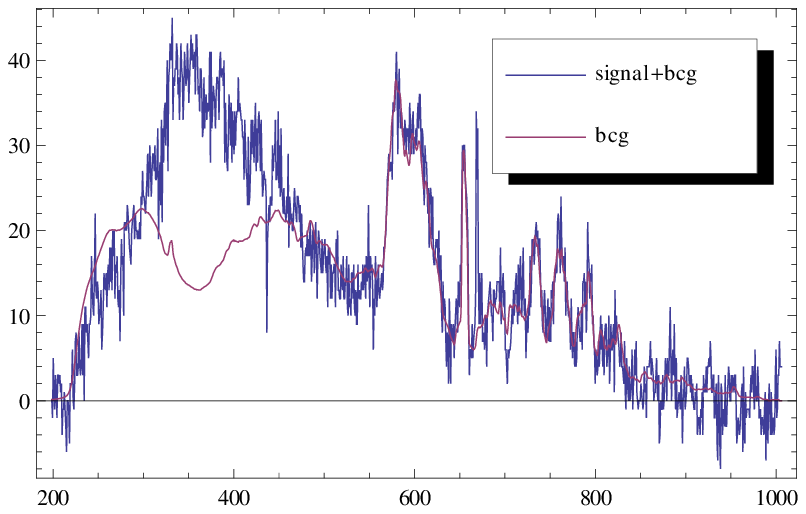} &
\includegraphics[width=2.5in]{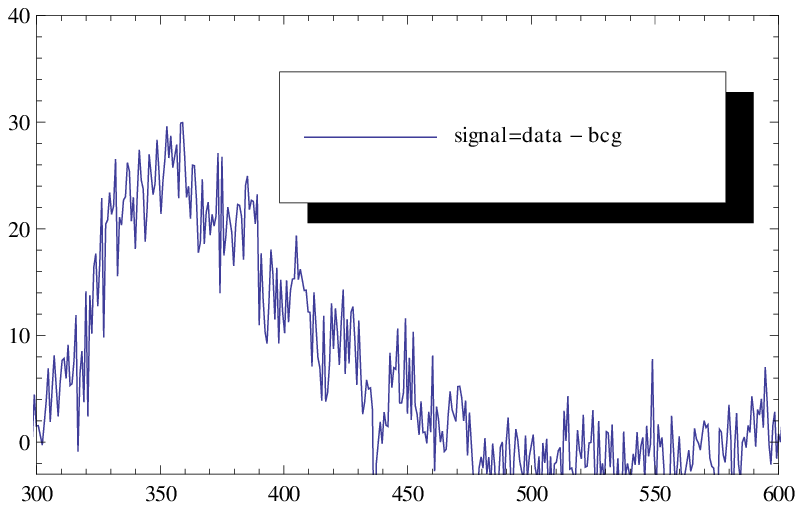} \\
\includegraphics[width=2.5in]{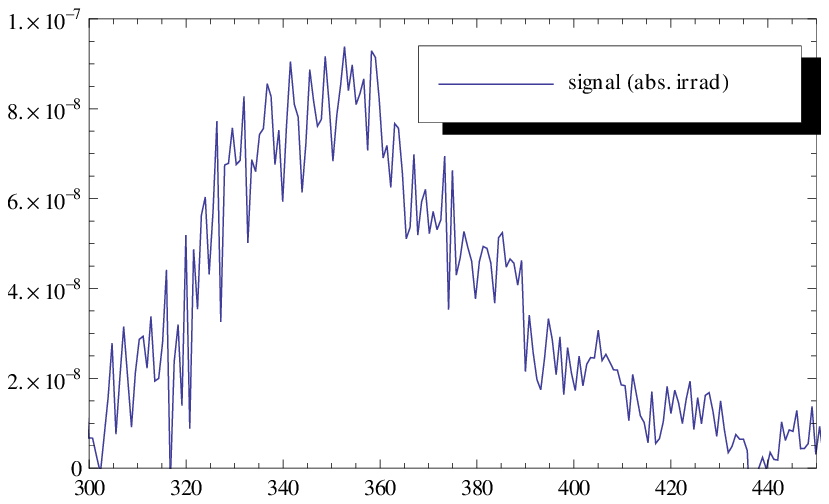} &
\includegraphics[width=2.5in]{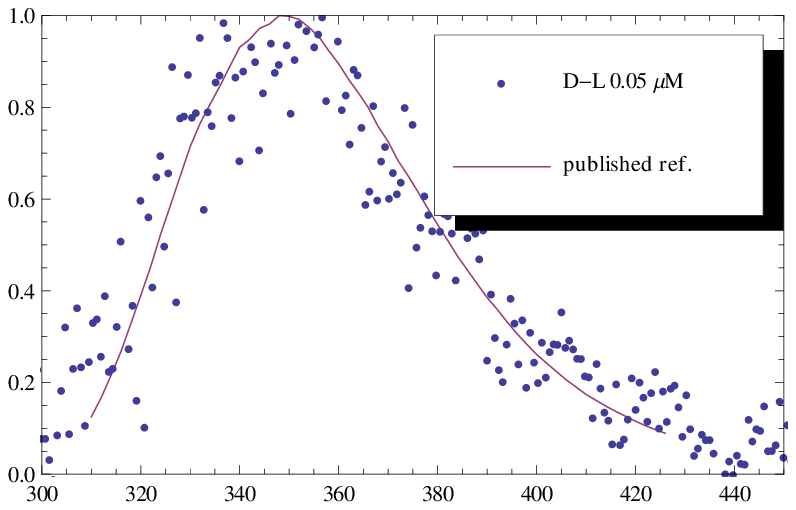} \\

\end{array}$
\end{center}
\caption{ Emission spectra study from D-L tryptophan of concentration 0.05 $\mu M$ in water. 
} 
\label{dl05}
\end{figure}

 Signal of fluorescence emission was detected
 at 90 degrees to beam-line by spectrometer (position I Fig. \ref{testcal}). Background from deuterium lamp is determined by measurements with spectrometer connected to
 connector on beam-line (position II Fig. \ref{testcal}). Spectrometer is switching between positions I and II.\\
 For normalization of background was used peak $D_{\alpha}$ at 656 nm in signal and background spectrum. This peak in signal spectrum 
 is fully accounted for by background light from deuterium light source. It's position is well established and therefore is natural point to
 calibrate amount of background in spectrum dominated by signal. \\
  Integration time in case of signal measurements was 60s, in case of background spectrum it was 0.4 s.\\
\paragraph{D-L tryptophan concentration 0.05 $\mu M$ in water} \mbox{} \\
 Results of measurements for D-L tryptophan 0.05 $\mu M$ concentration sample is summarized in Fig. \ref{dl05}. Upper left plot represents
 3 independent measurements (raw dark current subtracted). They are pretty much consistent between each other. Large fraction of Fulcher $\alpha$ band and $D_{\alpha}$ peak
 are evidence of large fraction of background at this conditions. Background shape is displayed in upper right plot. Middle left plot
 represents signal data spectra and estimated background, middle right ``pure'' signal spectrum - background subtracted from signal data 
 spectrum. Bottom left is calibrated spectrum using SL calibration correction. Bottom right is spectrum normalized to maximal value and compared
 with reference published measurements \cite{gard}. Published results are pretty much consistent with our final distribution.

\begin{figure}[h]
\begin{center}$
\begin{array}{cc}
\includegraphics[width=2.5in]{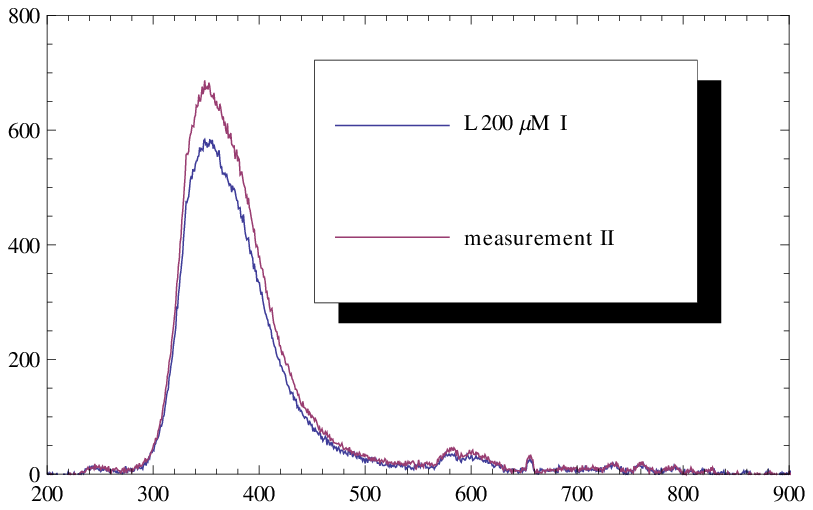} &
\includegraphics[width=2.5in]{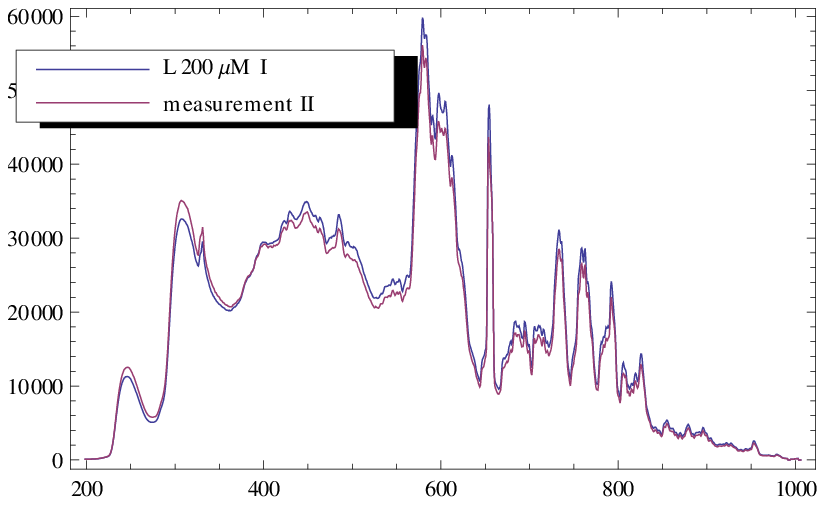} \\
\includegraphics[width=2.5in]{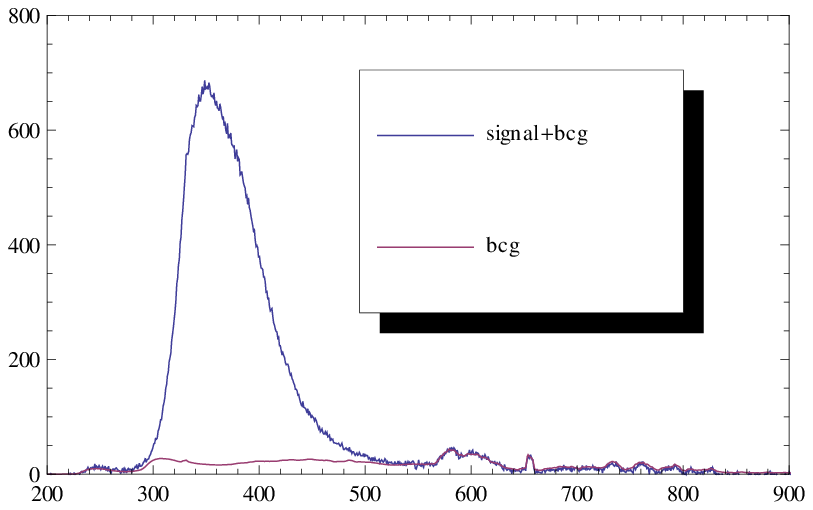} &
\includegraphics[width=2.5in]{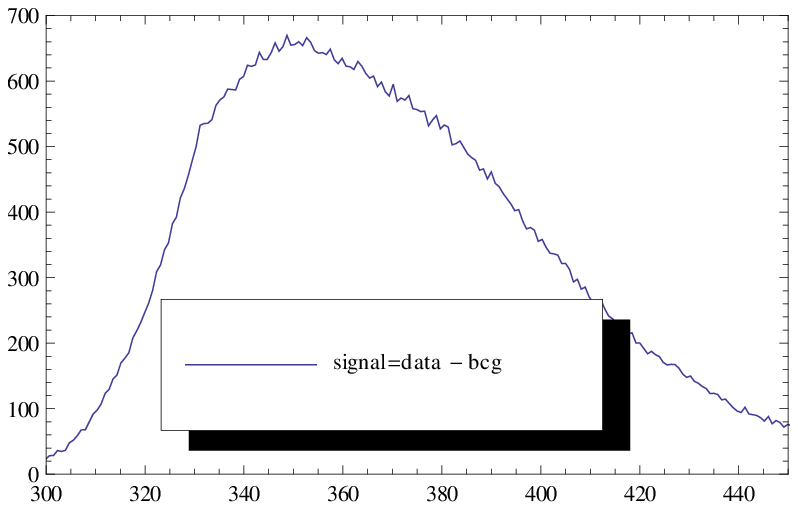} \\
\includegraphics[width=2.5in]{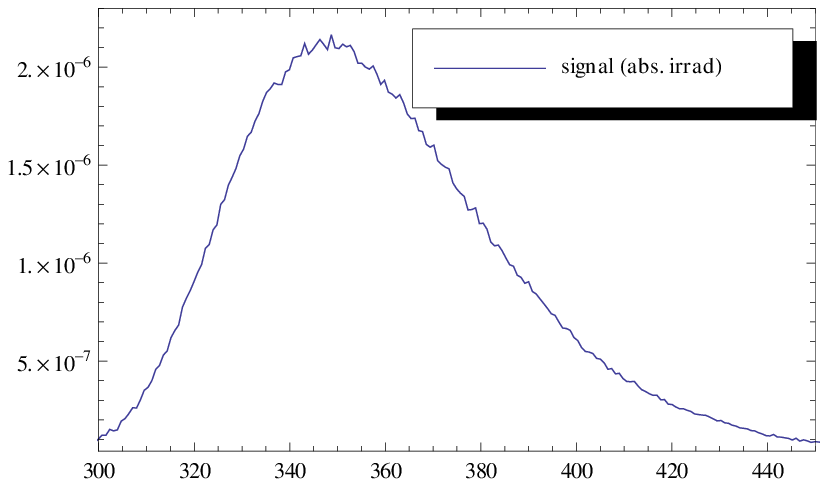} &
\includegraphics[width=2.5in]{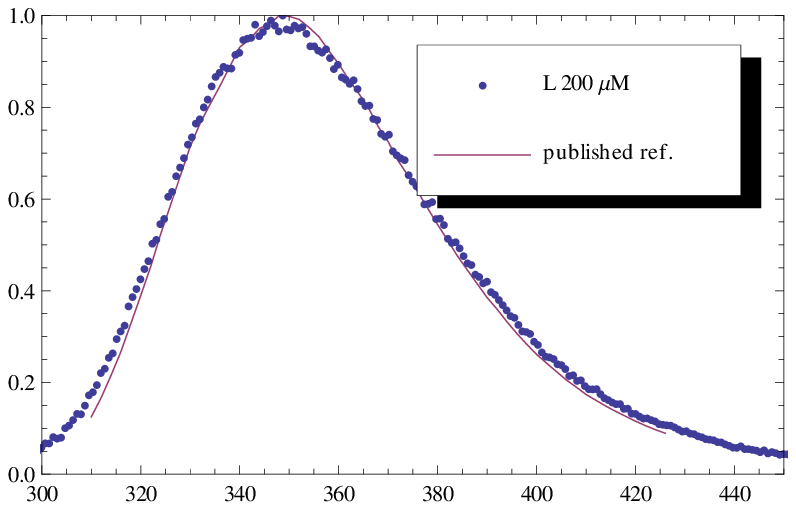} \\

\end{array}$
\end{center}
\caption{ Emission spectra study from L tryptophan of concentration 200 $\mu M$ in water. 
} 
\label{l200}
\end{figure}

\clearpage
\paragraph{L-tryptophan concentration 200 $\mu M$ in water} \mbox{} \\
 Same procedure is repeated for L-tryptophan of concentration 200 $\mu M$ in water. Experimental setup is the same. Results are presented
in Fig. \ref{l200}. Meaning of different plots is the same as in Fig. \ref{dl05}.\\
One can see that signal is very strong and background quite negligible. Final comparison between reference and our measurement is in 
very good agreement. 

\begin{figure}[h]
\begin{center}$
\begin{array}{cc}
\includegraphics[width=2.5in]{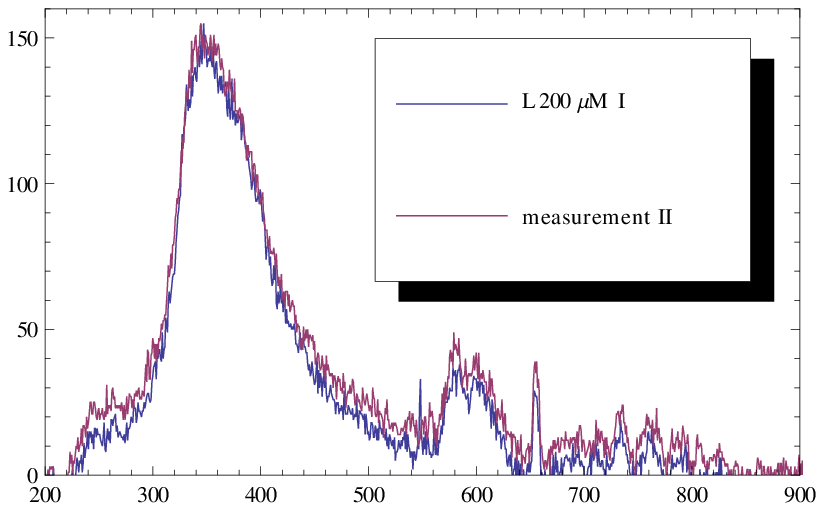} &
\includegraphics[width=2.5in]{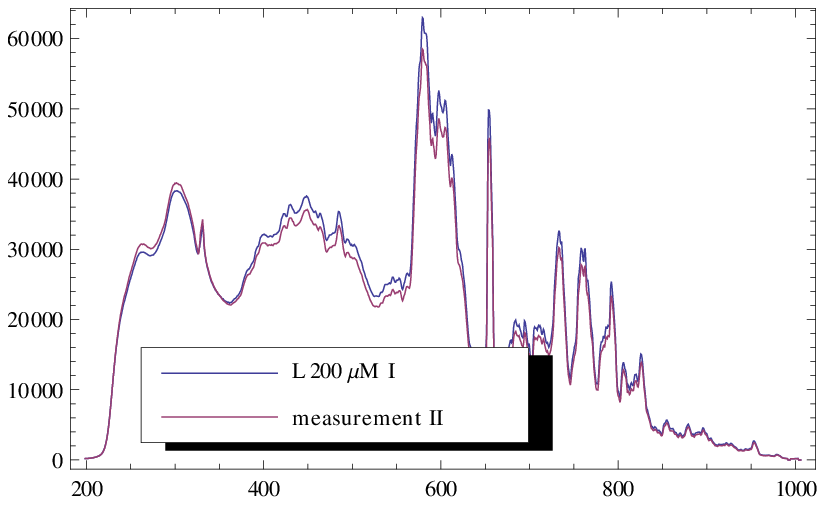} \\
\includegraphics[width=2.5in]{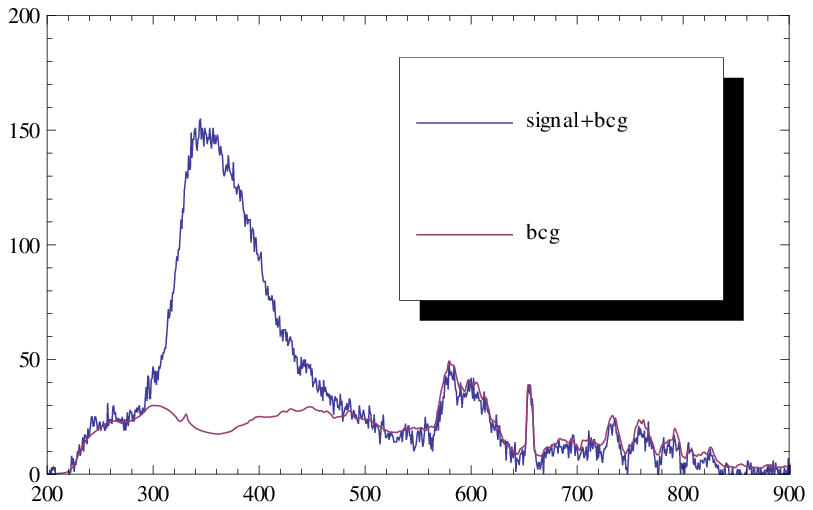} &
\includegraphics[width=2.5in]{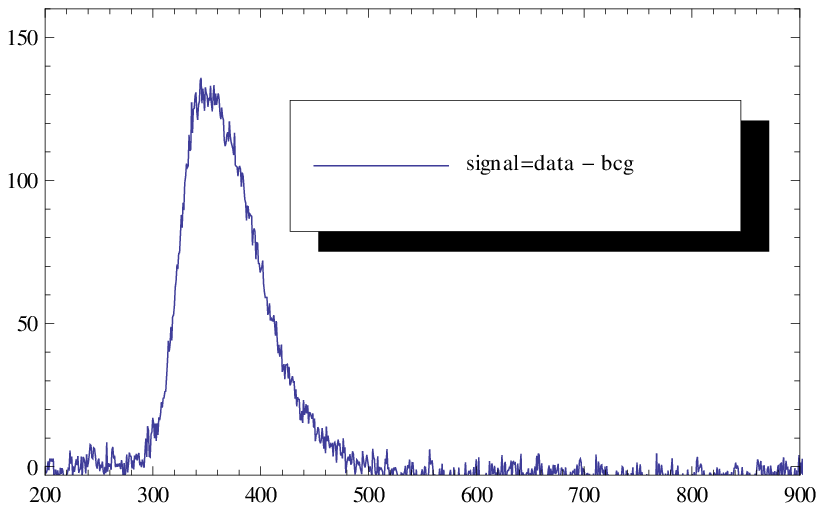} \\
\includegraphics[width=2.5in]{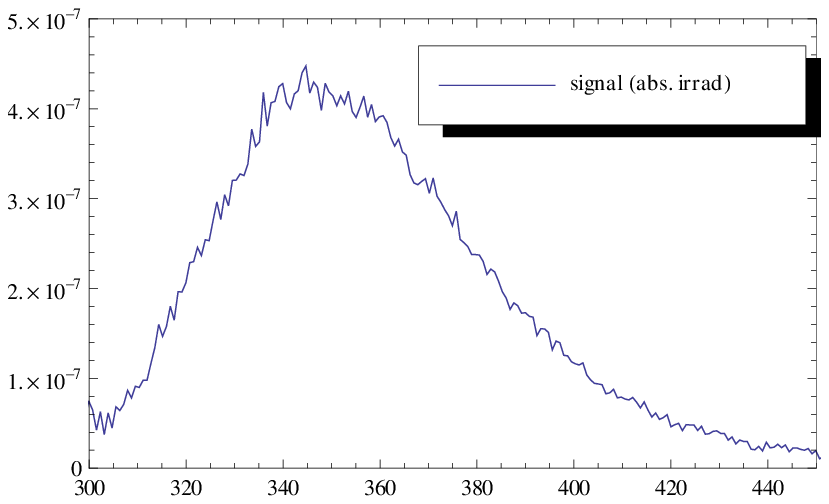} &
\includegraphics[width=2.5in]{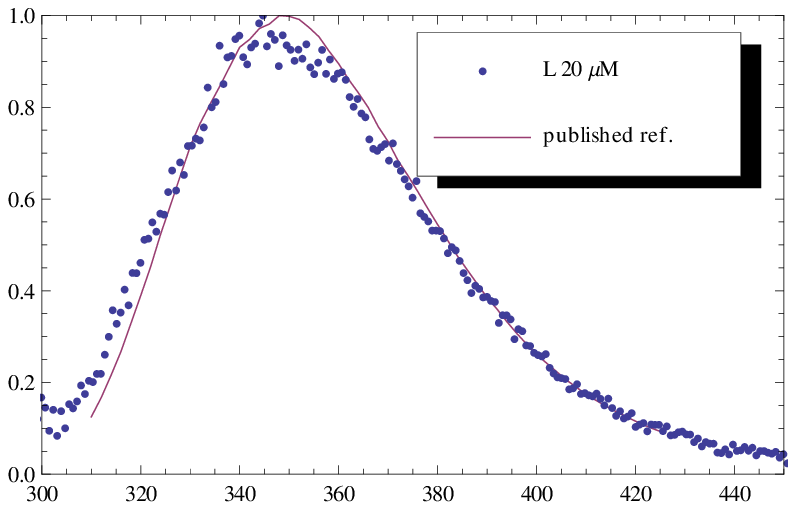} \\

\end{array}$
\end{center}
\caption{ Emission spectra study from L tryptophan of concentration 20 $\mu M$ in water. 
} 
\label{l20}
\end{figure}

\clearpage
\paragraph{L-tryptophan concentration 20 $\mu M$ in water} \mbox{} \\
 Exercise is once again repeated for a concentration of 20 $\mu M$ L-tryptophan in water. Results are summarized in Fig. \ref{l20}. Signal is
 less strong than in previous case and amount of background is increased. Conclusion about reference and our measurement remains the same.
  We can conclude that SL calibration procedure works reasonably well.
\clearpage
\section{Conclusion}
  Method for in situ calibration of spectrometer by using sonoluminesce have been proposed and tested. In situ calibration has an advantage 
  in case of further study of sonoluminescence by the same apparatus that calibrated is full optical setup. \\
  Procedure described as a test of calibration (by using tryptophan) by using SBSL can be used by itself as a 
  calibration procedure. Tryptopham in combination with simple (uncalibrated) deuterium lamp (as described) represents  robust and stable conditions
  for calibration of spectrometer. 
 
\end{document}